\documentstyle[preprint,aps,eqsecnum]{revtex}

\begin{document}

{\tighten
\preprint{\vbox{\hbox{CALT--68--1933}\hbox{FERMILAB--PUB--94--106--T}
\hbox{JHU--TIPAC--940005}\hbox{NSF--ITP--94--43}\hbox{UCSD/PTH  
94--05}}}

\title{Inconclusive Inclusive Nonleptonic $B$ Decays}

\author{Adam F.~Falk\thanks{On leave from The Johns Hopkins  
University, Baltimore, Maryland}}
\address{ Department of Physics, University of California,~San Diego,   
La Jolla, California 92093}
\author{Mark B.~Wise}
\address{California Institute of Technology, Pasadena, California  
91125}
\author{Isard Dunietz}
\address{Fermi National Accelerator Laboratory, P.O.~Box 500,
Batavia, Illinois 60510}

\date{May 20, 1994}

\maketitle
\begin{abstract}
We reconsider the conflict between recent calculations of the  
semileptonic branching ratio of the $B$ meson and the experimentally
measured rate. Such calculations depend crucially on the application of
``local duality'' in nonleptonic decays, and we discuss the relation of
this assumption to the weaker assumptions required to compute the
semileptonic decay rate. We suggest that the discrepancy between theory
and experiment might be due to the channel with two charm quarks in the
final state, either because of a small value for $m_c$ or because of a
failure of local duality.  We examine the experimental consequences of
such solutions for the charm multiplicity in $B$ decays.
\end{abstract}

}

\section{Introduction}

Because of the large energy which is released, the decay of a heavy  
quark is essentially a short distance process.  This simple observation
has led to much recent progress in the calculation of the inclusive
decays of hadrons containing a heavy quark
\cite{CGG,SVsemi,MW,Mannel,tau,SVrare,FLS}.  The method relies on the
construction of a systematic expansion in the inverse of the energy
release, given approximately by the heavy quark mass, and hence works most
reliably in the bottom system.  In fact, it is expected that certain
features of inclusive bottom hadron decays may be reliably predicted with
the accuracy of a few percent.

Considerable attention has been paid to inclusive semileptonic
\cite{SVsemi,MW,Mannel,tau} and rare \cite{SVrare,FLS} $B$ decays, both
to total rates and to lepton and photon energy   spectra. There is little
controversy that these calculations rest on a firm theoretical
foundation.  However, it has been suggested to extend these methods to
include nonleptonic decays as well \cite{SVrare,BS}.  This proposal has
led to an intriguing conflict with experiment, as the predicted
nonleptonic widths differ significantly from those which may be extracted
from the measured semileptonic branching ratio of the $B$ meson
\cite{SVbaff}.  In this calculation, the short-distance expansion has  
been carried out to third order in the inverse mass $1/m_b$, and a  
reasonable analysis leads the authors of Ref.~\cite{SVbaff} to the
conclusion that it would be unnatural to find the source of the
discrepancy in uncalculated terms of higher dimension or higher order in
$\alpha_s$.

It is the purpose of this article to reconsider this problem, in
particular the assumptions on which the computation is based.  In
Section~II, we review the techniques used to treat inclusive decay rates,
with an eye to emphasizing the differences between the theoretical
foundations underlying the calculations of semileptonic and nonleptonic
decays.  In Section~III, we discuss the possible discrepancy between
theory and experiment in the $B$ semileptonic branching ratio. This might
be resolved by an unusually small value for $m_c$, or might involve the
failure of the key assumption, ``local duality'', underlying the
calculation of the nonleptonic rate.  In either case the enhancement of
decays into final states with two charm quarks is a likely consequence. 
In Section~IV we examine the implications of this for the charm
multiplicity in $B$ decays, for which present data do not seem to support
an enhancement resulting from the $b\to c\bar cs$ process. The unusual
feature of the data on inclusive $B$ decays is neither the semileptonic
branching ratio alone, nor the charm multiplicity alone, but rather the
combination of the two.  Brief concluding remarks are given in Section~V.

\section{Theoretical techniques}

The weak decay of $b$ quarks is mediated by operators of the form
\begin{equation}
   {\cal O} = J^\mu_h J_{\ell\mu}\,,
\end{equation}
where
\begin{eqnarray}
   J^\mu_h &=& \overline q \gamma^\mu(1-\gamma^5) b\,,\cr
   J^\mu_\ell &=& \overline q_1 \gamma^\mu(1-\gamma^5) q_2\qquad
   \text{or}\qquad \overline \ell \gamma^\mu(1-\gamma^5) \nu_\ell
\end{eqnarray}
are fermion bilinears.  The inclusive decay rate is given by a sum over
all possible final states $X$ with the correct quantum numbers,
\begin{equation}\label{gam}
   \Gamma\sim\sum_X \langle B|\,{\cal O}^\dagger\,|X\rangle
                    \langle X|\,{\cal O}\,|B\rangle\,.
\end{equation}
In this article we adopt the notation that a generic $B$ meson contains a
$b$ quark, rather than a $\bar b$ quark.  The optical theorem may be used
to rewrite Eq.~(\ref{gam}) as the imaginary part of a forward scattering
amplitude,
\begin{equation}\label{Tdef}
   \Gamma\sim\text{Im}\,\langle B|\,T\left\{{\cal O}^\dagger,{\cal  
   O}\right\}\,|B\rangle\,.
\end{equation}
One then would like to use perturbative QCD to extract information about
the time-ordered product appearing in Eq.~(\ref{Tdef}).  The extent to
which this is possible is precisely the extent to which inclusive decay
rates may be calculated reliably.

In the case of semileptonic decays, one may follow a systematic procedure
to justify the application of perturbative QCD \cite{CGG}.  Up to
negligible corrections of order $\alpha_{EM}$ and $G_F$, one may
factorize the matrix element of the four-fermion operator,
\begin{equation}\label{factorize}
   \langle X\,\ell(p_\ell)\bar\nu(p_{\bar\nu})|\,
   J^\mu_h J_{\ell\mu}\,|B\rangle = \langle X|\,J^\mu_h\,|B\rangle
   \langle \ell(p_\ell)\bar\nu(p_{\bar\nu})|\,J_{\ell\mu}
   |0\rangle\,\,,
\end{equation}
and consider only the time-ordered product of the quark currents.  One
then finds an expression in which the integral over the momenta of the
leptons is explict,
\begin{equation}\label{lepint}
   \Gamma\sim\int\text{d}y\,\text{d}v\cdot\hat q\,\text{d}\hat q^2
   \,L_{\mu\nu}(v\cdot\hat q,\hat q^2,y)\, W^{\mu\nu}(v\cdot\hat q,\hat
   q^2)\, ,
\end{equation}
where $L_{\mu\nu}$ is the lepton tensor and $W^{\mu\nu}$ the hadron  
tensor. Here the momentum of the external $b$ quark is written as  
$p_b^\mu=m_b v^\mu$. The other independent kinematic variables are
$q^\mu=p^\mu_\ell+p^\mu_{\bar\nu}$ and $y=2E_\ell/m_b$.  It is convenient
to scale all momenta by $m_b$, so $\hat q=q/m_b$.  The hadronic tensor
is given by
\begin{eqnarray}
   W^{\mu\nu} &=& \sum_X \langle B|\,J_h^{\mu\dagger}\,|X\rangle
                  \langle X|\,J_h^\nu\,|B\rangle \cr
              &=& -2\text{Im}\,\langle B|\,i\int\text{d}x\,e^{iq\cdot x}
                  \,T\left\{J_h^{\mu\dagger}(x),J_h^\nu(0)\right\}
                  \,|B\rangle\cr
              &\equiv& -2\text{Im}\,T^{\mu\nu}\,.
\end{eqnarray}
One may perform the integrals in $y$, $v\cdot\hat q$ and $\hat q^2$ in
Eq.~(\ref{lepint}) to compute the total semileptonic decay rate, or  
leave some of them unintegrated to obtain various differential  
distributions.

The doubly differential distribution $\text{d}\Gamma/\text{d}y\,
\text{d}\hat q^2$ is a useful case to consider.  Here we must perform  
the integral over $v\cdot\hat q$, for $y$ and $\hat q^2$ fixed. The
range of integration for $v\cdot\hat q$ is given by $(y+\hat
q^2/y)/2\le v\cdot\hat q\le (1+\hat q^2-\hat m_q^2)/2$, where $m_q$ is
the mass of the quark to which the $b$ decays, and $\hat m_q=m_q/m_b$.
This integration is pictured in Fig.~\ref{vdotqfig}a, along with the
analytic structure of $T^{\mu\nu}$ in the $v\cdot\hat q$ plane
\cite{CGG,CR}.\footnote{The discussion of the analytic structure of 
$T^{\mu\nu}$ given in Ref.~\cite{MW} is erroneous.  We thank
B.~Grinstein and A.I.~Vainshtein for discussions of this point.}  The
absence of a cut along the real axis in the region 
$(1+\hat q^2-\hat m_q^2)/2< v\cdot\hat q<((2+\hat m_q)^2-\hat q^2-1)/2$ is simple to
understand in terms of the invariant mass $p_H$ of the intermediate
hadronic state.  Such a state may contain no $b$ quarks, in which case it
is subject to the restriction $p_H^2 = (m_b v-q)^2\ge m_q^2$ (the
left-hand cut), or it may contain $bb\bar q$, in which case $p_H^2 =  
(m_b v+q)^2\ge (2m_b+m_q)^2$ (the right-hand cut).  Except in the limit
$\hat q^2=\hat q^2_{\text{max}}=(1-\hat m_q)^2$ and $m_q\to0$, the two
cuts do not pinch.

In Fig.~\ref{vdotqfig}a, we have already included only the imaginary  
part of $T^{\mu\nu}$ by integrating over the top of the cut and then back
underneath it.  In general, $T^{\mu\nu}$ along the physical cut will
depend on  $v\cdot\hat q$ in a complicated nonperturbative way.  We do
not necessarily know how to compute in QCD in the physical region where  
there are threshold effects. However, we may use Cauchy's Theorem to  
deform the contour of integration until it lies away from the cut
everywhere except at its endpoints, as illustrated in
Fig.~\ref{vdotqfig}b.  Along the new contour, we are far from the
physical region, and we may perform an operator product expansion for
$T^{\mu\nu}$ in perturbative QCD.  Only far from any physical
intermediate states is such a calculation necessarily valid. However,
this is enough to allow us to compute reliably certain smooth integrals
of $T^{\mu\nu}$ by deforming the contour of integration into the
unphysical region.  That we can compute integrals of $T^{\mu\nu}$ in
perturbation theory in this way is the property of ``global duality''.

Unfortunately, the contour in Fig.~\ref{vdotqfig} must still approach the
physical cut near the endpoints of the integration.  This introduces an
uncertainty into the calculation which cannot be avoided.  Still, one has
two arguments that this uncertainty is likely to be small.  First, for
large $m_b$, the portion of the contour which is within
$\Lambda_{\text{QCD}}$ of the physical cut scales as
$\Lambda_{\text{QCD}}/m_b$ and thus makes a small contribution to the
total integral.  Second, if the energy release into the intermediate
hadronic system is large compared to $\Lambda_{\text{QCD}}$, it is
reasonable to expect that $T^{\mu\nu}$ will be well approximated by
perturbative QCD even in the physical region.  This is because in   this
region the cut is dominated by multiparticle states, and hence the
strength of the imaginary part of $T^{\mu\nu}$ is a relatively smooth  
function of the energy.  While new thresholds associated with the
production of additional pions are found along the cut even in this
region, their effect is small compared to the smooth background of  
states to which they are being added.

This intuition, that for large enough energies one may perform the
operator product expansion directly in the physical region, is ``local
duality''.  While it is a reasonable property for QCD to have, it is
obviously a stronger assumption than that of global duality.  In
particular, it cannot be justified by analytic continuation into the
complex plane.  Rather, it rests on one's sense of how QCD ought to  
behave at high energies.  It is clear, as well, that the energy at which  
local duality takes effect will depend on the operators which appear in
the time-ordered product.  Hence the fact that local duality appears to  
work at a given energy in one process, such as in electron-positron
annihilation into hadrons, may be suggestive but does not prove that it
should hold at the same energy in another process.

To compute the inclusive semileptonic decay rate, then, one may use  
global duality except in a region along the contour of order
$\Lambda_{\text{QCD}}/m_b$, where one must approach the physical cut.   
In this small region one must resort to local duality to justify the  
operator product expansion.

Let us now turn to inclusive nonleptonic decays.  Here there is no
analogue of the factorization (\ref{factorize}) which we had in the
semileptonic case.  Hence there is no ``external'' momentum $q$ in which
one may deform the contour away from the physical region, leaving one
unable to use global duality in the transition to perturbative QCD.  In
this case, one is forced to invoke local duality from the outset if one
is to argue that the the time-ordered product $T^{\mu\nu}$ is computable. 
This clearly puts the calculation of inclusive nonleptonic $B$ decays on
a   less secure theoretical foundation than that of inclusive semileptonic
$B$ decays.

Nonetheless, we do not mean to assert that the assumption of local  
duality in nonleptonic decays is inherently unreasonable, merely that it
is the least reliable aspect of the computation.  In fact, it is not
entirely clear what it is reasonable to expect in this case.  On the one
hand, the energy released when a $b$ quark decays is certainly large
compared to $\Lambda_{\text{QCD}}$.  On the other, the decay is initially
into three strongly interacting particles (rather than into only one for
semileptonic decays), and the energy per strongly interacting particle is
not really so large.  (Note that in the semileptonic case, the point at
which the contour approaches the cut and local duality must be invoked is
conveniently the point of {\it maximum\/} recoil of the final state
quark, where local duality is expected to work best.)  What we propose is
that the comparison of the nonleptonic decay rate, as computed via the
operator product expansion, with experiment be taken as a direct test of
local duality in this process.  As such, it is a probe of a property of
QCD in an interesting kinematic region, and nonleptonic $B$ decay well
deserves the intense scrutiny which it has recently been accorded.

\section{The semileptonic branching fraction of $B$ mesons}

The experimental implications of inclusive nonleptonic decays of $B$
mesons have recently been discussed in great detail by Bigi, Blok,  
Shifman and Vainshtein \cite{SVbaff}.  Since the semileptonic branching
ratio of the $B$ is relatively well-measured, they use their calculation
of the nonleptonic decay rate to predict this quantity.  Their conclusion
is  that the semileptonic branching ratio which comes out of their
computation is unacceptably high, corresponding to a nonleptonic width
which is too low by at least $15$-$20\%$.  In this section we will
reconsider their analysis.

The inclusive decay rate of the $B$ meson may be divided into parts  
based on the flavor quantum numbers of the final state,
\begin{equation}\label{totalrate}
   \Gamma_{\text{TOT}} = \Gamma(b\to c\,\ell\bar\nu) + \Gamma(b\to
   c\bar ud') + \Gamma(b\to c\bar cs')\,.
\end{equation}
Here we neglect rare processes, such as those mediated by an underlying
$b\to u$ transition or penguin-induced decays.  By $d'$ and $s'$ we mean
the approximate flavor eigenstates $(d'=d\cos\theta_1-s\sin\theta_1$,
$s'=d\sin\theta_1+s\cos\theta_1)$ which couple to $u$ and $c$,
respectively, and we ignore the effect of the strange quark mass.  It is
convenient to normalize the inclusive partial rates to the  
semielectronic rate, defining
\begin{equation}
   R_{ud} = {\Gamma(b\to c\bar ud')\over
             3\Gamma(b\to c\,\text{e}\bar\nu)}\,,\qquad\qquad
   R_{cs} = {\Gamma(b\to c\bar cs')\over
             3\Gamma(b\to c\,\text{e}\bar\nu)}\,.
\end{equation}
The full semileptonic width may be written in terms of the  
semielectronic width as
\begin{equation}
   \Gamma(b\to c\,\ell\bar\nu) = 3f(\hat m_\tau)
   \Gamma(b\to c\,\text{e}\bar\nu)\,,
\end{equation}
where the factor $3f(\hat m_\tau)$ accounts for the three flavors of
lepton, with a phase space suppression which takes into account the  
$\tau$ mass.  Then, since the semileptonic branching ratio is given by
$Br(b\to c\,\ell\bar\nu) = \Gamma(b\to c\,\ell\bar\nu)/
\Gamma_{\text{TOT}}$, we may rewrite Eq.~(\ref{totalrate}) in the   form
\begin{equation}\label{rsum}
   R_{ud} + R_{cs} = f(\hat m_\tau)\,{1-Br(b\to c\,\ell\bar\nu)
   \over Br(b\to c\,\ell\bar\nu)}\,.
\end{equation}
The measured partial semileptonic branching fractions are \cite{pdb,aleph}
\begin{eqnarray}
   Br(B\to X\text{e}\bar\nu) &=& 10.7\pm0.5\%\,,\cr
   Br(B\to X\mu\bar\nu) &=& 10.3\pm0.5\%\,,\cr
   Br(B\to X\tau\bar\nu) &=& 2.8\pm0.6\%\,,
\end{eqnarray}
leading to a total semileptonic branching fraction $Br(b \rightarrow
c\,\ell\bar\nu)$  of $23.8\pm0.9\%$, with the experimental errors added
in quadrature.  Of the semileptonic rate, $11\%$ comes from decays to
$\tau$, corresponding to a phase space suppression factor $f(\hat
m_\tau)=0.74$, consistent with what one would expect in free quark
decay\cite{tau,CPT}. If we substitute the measured branching fractions
into the right-hand   side of Eq.~(\ref{rsum}), we find
\begin{equation}\label{rsumdata}
   R_{ud} + R_{cs} = 2.37\pm0.12\,.
\end{equation}
We now compare this constraint with the theoretical calculations of
$R_{ud}$ and $R_{cs}$.

The ratios $R_{ud}$ and $R_{cs}$ depend on the total rates  
$\Gamma(b\to c\,\text{e}\bar\nu)$, $\Gamma(b\to c\bar ud')$ and
$\Gamma(b\to c\bar cs')$.  Each of these has a theoretical expansion in
terms of $\alpha_s(\mu)$ and $1/m_b$.  Since corrections of order $1/m_b$
vanish and those of order $1/m_b^2$ are numerically expected to be at the
few percent level \cite{CGG,SVsemi,MW,Mannel,SVrare,FLS,SVbaff}, we
include here only the radiative corrections.  Neglecting terms of order
$\alpha_s^2(\mu)$, the expansions take the form
\begin{eqnarray}\label{gammas}
   \Gamma(b\to c\,\text{e}\bar\nu) &=& \Gamma_0 I(\hat m_c,0)
   \cdot\left\{ 1 - {2\alpha_s(\mu)\over3\pi}\left(\pi^2-{25\over4}+
   \delta_{\text{s.l.}}(\hat m_c)\right)\right\}\,,\cr
   \Gamma(b\to c\bar ud') &=& \Gamma_0 I(\hat m_c,0)\cdot
   3\eta(\mu)\left\{ 1 -  
  {2\alpha_s(\mu)\over3\pi}\left(\pi^2-{31\over4}+
   \delta_{ud}(\hat m_c)\right) + J_2(\mu) \right\}\,,\cr
   \Gamma(b\to c\bar cs') &=& \Gamma_0 I(\hat m_c,0)\cdot
   3\eta(\mu) G(\hat m_c)
   \left\{ 1 - {2\alpha_s(\mu)\over3\pi}\left(\pi^2-{31\over4}+
   \delta_{cs}(\hat m_c)\right) + J_2(\mu) \right\}\,.
\end{eqnarray}
The prefactor $\Gamma_0=G_F^2m_b^5|V_{cb}|^2/192\pi^3$ will cancel in  
the ratios $R_{ud}$ and $R_{cs}$, as will the charm quark phase space
suppression $I(\hat m_c,0)$ \cite{CPT}, to be discussed below.

The radiative corrections have been computed analytically to order
$\alpha_s$ in the limit $m_c=0$, and for semileptonic decays up to   one
numerical integration for general $m_c$ \cite{JK}.  For semileptonic
decays we absorb the correction due to $\hat m_c\ne 0$ into
$\delta_{\text{s.l.}} (\hat m_c)$ and present the numerical value of
$\delta_{\text{s.l.}} (\hat m_c)$ below.  Finite charm mass effects for
nonleptonic decays are absorbed into $\delta_{ud}(\hat m_c)$ and
$\delta_{cs} (\hat m_c)$.  Because these quantities have not been
computed, we present numerical results in the case of nonleptonic  
decays only for $\hat m_c = 0$. The expressions for $\Gamma(b\to c\bar
ud')$ and $\Gamma(b\to c\bar cs')$ in Eq.~(\ref{gammas}) are somewhat more
complicated than that for $\Gamma(b\to c\,\text{e}\bar\nu)$, due to
renormalization group running between $\mu=M_W$ and $\mu=m_b$.  The
leading logarithms are resummed into $\eta=(2L_+^2+L_-^2)/3$, where
\cite{GLAM}
\begin{equation}
   L_+=\left[{\alpha_s(\mu)\over\alpha_s(M_W)}\right]^{-6/23}\,,
   \qquad\qquad
   L_-=\left[{\alpha_s(\mu)\over\alpha_s(M_W)}\right]^{12/23}\,.
\end{equation}
The subleading logarithms, which must be included if terms of order
$\alpha_s(\mu)$ are also to be kept, are assembled into $J_2$,
\begin{equation}
   J_2 = {2\alpha_s(\mu)\over3\pi}\left({19\over4}+6\log{\mu\over m_b}
         \right){L_-^2-L_+^2\over2L_+^2+L_-^2}
         + 2\left({\alpha_s(\mu)-\alpha_s(M_W)\over\pi}\right)
         {2L_+^2\rho_++L_-^2\rho_-\over2L_+^2+L_-^2}\,,
\end{equation}
where $\rho_+=-{6473\over12696}=-0.51$ and $\rho_-={9371\over6348}=1.47$
arise from two-loop anomalous dimensions~\cite{AP}.  The factor 3 in  
Eq.~(\ref{gammas}) is for the sum over colors in the final state. 
Finally, there is an additional phase space suppression $G(\hat m_c)$ in
$\Gamma(b\to c\bar cs')$ because of the masses of the two charm quarks. 
This factor is given by \cite{CPT}
\begin{equation}
   G(\hat m_c) = {I(\hat m_c,\hat m_c)\over I(\hat m_c,0)}\,,
\end{equation}
where
\begin{eqnarray}
   I(x,0) &=& (1-x^4)(1-8x^2+x^4)-24x^4\log x\,,\cr
   I(x,x) &=& \sqrt{1-4x^2}(1-14x^2-2x^4-12x^6)+24x^4(1-x^4)               
   \log\left({1+\sqrt{1-4x^2}\over1-\sqrt{1-4x^2}}\right)\,.
\end{eqnarray}
In terms of the theoretical expressions (\ref{gammas}) for the partial
widths, the ratios take the form
\begin{eqnarray}\label{rbars}
   R_{ud} &=& P(\mu)+\delta P_{ud}(\mu,\hat m_c)\,,\cr
   R_{cs} &=& G(\hat m_c)\bigg[P(\mu)+\delta P_{cs}(\mu,\hat
              m_c)\bigg]\,,
\end{eqnarray}
where
\begin{eqnarray}
   P(\mu) &=&\eta(\mu)\left[1+{\alpha_s(\mu)\over\pi}
             +J_2(\mu)\right]\,,\cr
   \delta P_{ud}(\mu,\hat m_c) &=& \eta(\mu){2\alpha_s(\mu)\over3\pi}
   \bigg[\delta_{\text{s.l.}}(\hat m_c)-\delta_{ud}(\hat m_c)\bigg]\,,\cr
   \delta P_{cs}(\mu,\hat m_c) &=& \eta(\mu){2\alpha_s(\mu)\over3\pi}
   \bigg[\delta_{\text{s.l.}}(\hat m_c)-\delta_{cs}(\hat m_c)\bigg]
\end{eqnarray}
parametrize the radiative corrections.  As emphasized in
Ref.~\cite{SVbaff}, if the theoretical expressions (\ref{rbars}) are
inserted, then Eq.~(\ref{rsumdata}) is not well satisfied. For example, if
one simply takes the reasonable values $\mu=m_b=4.8\,\text{GeV}$,
$m_c=1.5\,\text{GeV}$,
$\Lambda^{(5)}_{\overline{\text{MS}}}=180\,\text{MeV}$ and
$\delta P_{ud}=\delta P_{cs}=0$, then $P(\mu) = 1.27$, $G(\hat m_c)=0.36$
and the left-hand side of Eq.~(\ref{rsumdata}) is only $1.73$.  We are
thus tempted to push the uncertainties in the calculation as far as is
reasonable, in order to see how much of the discrepancy can be resolved
within the context of the operator product expansion.

The largest uncertainty in the theoretical expression for  
$R_{ud}+R_{cs}$ comes from the choice of the charm and bottom masses.  Up
to certain ambiguities which have recently been
discussed~\cite{renormalons}, within perturbation theory these masses
should be taken to be the pole masses \cite{MW,GFLN}.  These masses have
not been determined with much precision.  However, within the heavy quark
expansion, the difference between $m_c$ and $m_b$ is much more precisely
known, in terms of the spin-averaged $D$ meson and $B$ meson masses:
\begin{equation}\label{massdiff}
   m_b-m_c=\langle M_B\rangle_{\text{ave.}} -
           \langle M_D\rangle_{\text{ave.}} = 3.34\,\text{GeV}\,.
\end{equation}
In what follows, we will hold $m_b-m_c$ fixed, and consider variations
of $m_b$ only.  A reasonably conservative range for $m_b$ might be
$4.4\,\text{GeV}\le m_b\le 5.0\,\text{GeV}$, which corresponds to
$0.24\le\hat m_c\le 0.33$.  In Fig.~\ref{Gfig}, we plot $G(\hat m_c)$  
as a function of $m_b$, using the constraint (\ref{massdiff}).  In
Fig.~\ref{Pfig}, we plot $P(\mu)$ for a variety of values of the QCD
scale $\Lambda^{(5)}_{\overline{\text{MS}}}$ \cite{pdb}.

We start by considering $R_{ud}$, for which the calculation is likely  
to be more reliable, since it is less sensitive to $\hat m_c$.  There is
uncertainty in the radiative correction $P(\mu)$ from the choice of the
renormalization scale $\mu$.  The usual choice $\mu=m_b$ is motivated by
the fact that the total energy released in the decay is $m_b$.  However,
this energy has to be divided between three particles, so perhaps the
appropriate scale is lower.  For
$\mu=1.6\,\text{GeV}\approx m_b/3$, a reasonable lower limit, and
$\Lambda^{(5)}_{\overline{\text{MS}}}=180\,\text{MeV}$, we find
$P(\mu)=1.45$, a modest enhancement over $\mu=4.8\,\text{GeV}$.  If
$\Lambda^{(5)}_{\overline{\text{MS}}}$ is taken as high as
$220\,\text{MeV}$, we have $P(\mu)=1.52$, which makes a small additional
difference.  The uncertainty in $\delta P_{ud}$ is harder to estimate,
since $\delta_{ud}(\hat m_c)$ has not been calculated.  However, one may
extract $\delta_{\text{s.l.}}(\hat m_c)$ by doing a numerical  
integration of the formulas in Ref.~\cite{JK}.  For $\hat m_c=0.30$, we
find $\delta_{\text{s.l.}}=-1.11$, corresponding to 
$(2\alpha_s(m_b)/3\pi)\delta_{\text{s.l.}}(\hat m_c)=-0.050$.  The
magnitude of this correction grows approximately linearly with $\hat  
m_c$, and for $\hat m_c=0.33$, we have $\delta_{\text{s.l.}}=-1.20$. 
Hence the term is small and actually reduces $R_{ud}$, although one might
expect it to cancel in whole or in part against the term proportional to
$\delta_{ud}(\hat m_c)$.  What we can conclude at this point is that the
error associated with ignoring the charm quark mass in the radiative
corrections is likely to be no larger than $\pm0.05$, and henceforth we
will neglect this effect.

The leading nonperturbative strong interaction corrections to $R_{ud}$
and $R_{cs}$ are characterized by the two dimensionless quantities
$K_b=-\langle B(v)|\,\overline b_v (iD)^2 b_v\,|B(v)\rangle/2m_b^2$ and
$G_b = \langle B(v)|\,\overline b_v g_s G_{\mu\nu} \sigma^{\mu\nu}  
b_v\,|B(v)\rangle/4m_b^2$. Because it breaks the heavy quark spin
symmetry, the parameter $G_b$ may be determined from the measured $B^* -
B$ mass splitting, but the value of $K_b$ is not known.  Fortunately,
$K_b$ does not occur in the nonperturbative correction to $R_{ud}$.
(Using the ``smearing'' technique of Ref.~\cite{MW}, this cancelation
arises because $\Gamma (b\to c\bar ud)$ and $\Gamma (b\to
c\text{e}\bar\nu)$ have the same dependence on $m_b$.)  For $R_{ud}$,
then, we are more confident than for $R_{cs}$ that the nonperturbative
QCD corrections are small.  Note, however, that there is a contribution
to the mass difference in Eq.~(\ref{massdiff}) involving $K_b$ and $K_c$,
which we have neglected.

The above estimates lead us to the conclusion that with the effects we
have included in the operator product expansion, it is difficult to  
avoid the upper bound $R_{ud}\le1.52$.  If this is true, then
Eq.~(\ref{rsumdata}) would imply $R_{cs}\ge0.85$.  This can barely be
achieved in the theoretical expressions we have given.  If we vary
$4.4\,\text{GeV}\le m_b\le5.0\,\text{GeV}$, as suggested above, then
$0.27\le G(\hat m_c)\le0.58$.  Estimating the radiative corrections as
before, with $\Lambda^{(5)}_{\overline{\text{MS}}}=220\,\text{MeV}$,  
this suggests the upper limit $R_{cs}\le0.89$, or
$R_{ud}+R_{cs}\le2.43$.  This is in agreement with experiment, but on
the other hand, it requires us to push all the freedom in the
calculation in the same direction, perhaps further than is reasonable. 
If one were to take the point of view that
$\mu=2.4\,\text{GeV}\approx m_b/2$ were the lowest reasonable value for
$\mu$, then one would have the constraints $R_{ud}\le1.44$,  
$R_{cs}\le0.83$ and $R_{ud}+R_{cs}\le2.27$.  If one were further to
require $m_b\ge4.6\,\text{GeV}$, one would have $R_{cs}\le0.67$ and
$R_{ud}+R_{cs}\le2.10$.  In this case, one might consider the  
discrepancy with experiment to be a more serious issue.

Another possibility is that the relevant scale for the radiative
corrections in the decay to two charm quarks is considerably lower than
that for the final state with a single charm.  Since the rest masses of
the two charm quarks absorb approximately 60\% of the energy available
in the decay, the strongly interacting particles are not emitted with
very large momenta.  For example, the average energy of the strange
quark in the decay $b \rightarrow c\bar cs$, computed at tree level, is
only about $1\,\text{GeV}$.  With such a low energy the procedure of
estimating the value of higher order QCD corrections by varying the
subtraction point $\mu$ is of dubious value. In fact one might question
whether any finite order of perturbation theory is adequate and whether
threshold effects that cause a violation of local duality are important.

It is evident  from this discussion that nothing is particularly clear.
Although the data on inclusive nonleptonic decays can almost be accounted
for by squeezing the input parameters, one might feel a little nervous
about the necessity of such a conspiracy.  After all, as mentioned
earlier the ``reasonable'' values $\mu=m_b=4.8\,\text{GeV}$,
$m_c=1.5\,\text{GeV}$ and $\Lambda^{(5)}_{\overline{\text
{MS}}}=180\,\text{MeV}$ lead to $R_{ud}=1.27$ and $R_{cs}=0.46$, far short
of the mark.  An enhancement of approximately $40\%$ in the nonleptonic
rate is called for.  If one were to require this effect to be found
entirely in $R_{cs}$, it would amount to more than a factor of two. 
While we are less inclined than the authors of Ref.~\cite{SVbaff} to
insist that something is amiss, it is nonetheless intriguing to consider
the possibility that the data indicate an enhancement of the nonleptonic
rate over and above what we have included in the operator product
expansion. Where might such an enhancement come from?

The simplest explanation would be that due to a failure of local  
duality, the inclusive nonleptonic decay rate is simply not calculable
to better than $40\%$ or so.  This is certainly a discouraging
explanation, in that if it were true then there would be very little
one could say in detail about why local duality, and hence the
calculation, had failed. One was simply unlucky.  On the other hand,
this explanation may well be correct. While we expect local duality to
hold in the asymptotic limit of infinte $b$ quark mass, we have little to
guide us in estimating how heavy the $b$ quark actually needs to be in
practical terms.  In particular, it is not relevant to consider, at low
orders in QCD perturbation theory,  the size of a few subleading terms
which appear in the operator product expansion itself.  The matrix
elements which appear in  this expansion are sensitive to details of
the $B$ meson bound state, but they are explicitly not sensitive to
resonance effects in the final hadronic state. 

If local duality fails, it could well fail differently in the
$\Gamma(b\to c\bar ud')$ and $\Gamma(b\to c\bar cs')$ channels. In fact,
we would expect it to fail worse in the channel with two charm quarks,  
since we expect the final states to be characterized by lower particle
multiplicity and be closer to the resonance-dominated regime. Local
duality, by contrast, is applicable only in the regime where the effect
of individual resonance thresholds is small compared to the almost  
smooth ``continuum'' of multiparticle states.  On the other hand, the
phase space suppression from the two final state charm quarks means
that unless $m_c$ is unusually small, only thirty percent or so of the
inclusive nonleptonic rate comes from the $\Gamma(b\to c\bar cs')$
channel. Hence, to account for an enhancement of the full nonleptonic
rate by forty percent purely from $b\to c\bar cs'$ would require a
dramatic failure of local duality in this channel.

\section{Experimental consequences of an enhancement of  
$R_{\lowercase{cs}}$}

Either through a failure of local duality, or from an unusually small
value for $m_c$, or because of a combination of these effects, the  
value of $R_{cs}$ is likely to be near unity in order to account for the  
measured $B$ semileptonic branching ratio.  This corresponds to about
one-third of $B$ decays arising from the $b \to c\bar cs'$ process.  One
consequence of this is a large number of charmed quarks per $B$ decay,
\begin{equation}\label{nc}
   n_c = 1 + R_{cs}\, {Br (B \to X_c \ell\bar \nu)\over f (\hat
   m_\tau)}\,.
\end{equation}
We remind the reader that we have adopted the notation that a generic
$B$ meson contains a $b$ quark, rather than a $\bar b$ quark.  Using $Br
(B\to X_c\,\ell\bar\nu) = 23.8\%$ and $f(\hat m_\tau) = 0.74$ in
Eq.~(\ref{nc}) yields
\begin{equation}
   n_c = 1.00 + 0.32\,R_{cs}\,,
\end{equation}
which for the values of $R_{cs}$ necessary to explain the semileptonic
branching ratio would indicate $n_c\sim 1.3$.  

There are contributions to the experimental value of $n_c$ from charmed
mesons, charmed baryons, and $c\bar c$ resonances.  The number of charged
and neutral $D$ mesons per decay, summed over $B$ and $\overline B$, has
been measured to be~\cite{BHP}
\begin{eqnarray}\label{dmulti}
   n_{D^\pm} &=& 0.246\pm0.031\pm0.025\,,\cr
   n_{D^0,\overline D^0} &=& 0.567\pm0.040\pm0.023\,.
\end{eqnarray}
The branching ratio to $D_s^\pm$ mesons has not yet been determined,
because no absolute $D_s$ branching ratio has been measured.   
However, it is known that~\cite{menary}
\begin{equation}\label{dsdata}
   n_{D^{\pm}_s} = (0.1224 \pm0.0051\pm0.0089)
   \left[{3.7\%}\over Br (D_s\to\phi \pi)\right]\,,
\end{equation}
and the branching ratio for $D_s\to\phi\pi$ is expected to be about
$3.7\%$. 

We must include in $n_c$ twice the inclusive branching ratio to all $c\bar
c$ resonances which are below $D\overline D$ threshold.  The measured
inclusive branching ratio to $\psi$ is $(1.11\pm 0.08)\%$, including
feed-down from $\psi'$ and $\chi_c$ decays~\cite{BHP}. It is also known
that $Br(B\to\psi'X)=(0.32\pm0.05)\%$, $Br(B\to\chi_{c1}
X)=(0.66\pm0.20)\%$ and $Br(B\to\eta_c X)<1\%$. Hence we expect that the
inclusive B branching ratio to charmonium states below $D\overline D$
threshold is about $2\%$.

The inclusive $B$ decay rate to baryons is about $6\%$ \cite{BHP}.  While
it is commonly believed that these baryons arise predominantly from the
$b\to c\bar ud'$ process, giving $\Lambda_cX$ final states, we argue
elsewhere \cite{DFW} that a large fraction of $B$ decays to baryons
actually arise from the $b\to c\bar cs'$ process, which gives final
states with both a charm baryon and an anticharm baryon, such as $\Xi_c
\overline\Lambda_cX$. Evidence for this interpretation comes from the
experimental distribution of $\Lambda_c$ momenta, which shows that the  
$\Lambda_c$'s produced in $B$ or $\overline B$ decay are recoiling against
a state with a mass greater than or equal to the mass of the $\Xi_c$
\cite{al,cr,zo}. This novel interpretation of $B$ decays to baryons can  
be consistent with the measured $\Lambda\ell ^{\pm}$ correlations if 
$Br(\Xi_c \to \Lambda X)/Br(\Lambda_c \to \Lambda X)$ is
large~\cite{DFW}.

Even if $B$ decay to baryons predominantly gives final states with  
both a charm and an anticharm baryon, the data summarized above do not
provide supporting evidence for a value of $n_c$ around 1.3. Given the
uncertainties, however, such a large value for the number of charmed
hadrons per $B$ decay is perhaps not excluded. From our perspective the
curious feature of the data on inclusive $B$ decay is not the measured
semileptonic branching ratio alone, but rather the combination of it with
the data on charm multiplicity in these decays.

In this paper we have neglected $B$ decays that do not arise from an
underlying $b\to c$ transition. Other possible processes include the $b\to
u$ transition and contributions from penguin-type diagrams. While it is
very unlikely that such sources contribute significantly to the
nonleptonic decay rate, this assumption can be tested experimentally, if
enough branching ratios can be measured.  The fraction of $B$
decays arising from the $b\to c$ transition is given by the sum of the $B$
branching ratio to charmonium states below $D\overline D$ threshold, the
branching ratio to states containing at least one charmed baryon, and the
branching ratios to the ground state charmed mesons, $Br(B\to D^0X)$,
$Br(B\to D^+X)$ and $Br(B\to D_s^+X)$. Note that the inclusive charm
yields reported in Eqs.~(\ref{dmulti}) and (\ref{dsdata}) are actually
sums of branching ratios (for example, neglecting $CP$ violation,
$n_{D^{\pm}}=Br(B\to D^+X)+Br(B\to D^-X)$). However, it should be
possible with enough data to extract the individual branching ratios
themselves.  

For example, one could count the number of $DD$ (or $\overline
D\,\overline D$) events per
$B\overline B$ event at the $\Upsilon(4S)$,
\begin{eqnarray}\label{ndd}
   n(DD) &=&(1-b)\,Br(B\to DX)\,Br(\overline B\to DX) \cr
   &&+{b\over2}\left[Br(B\to DX)\,Br(B\to DX)+
   Br(\overline B\to DX)\,Br(\overline B\to DX)\right] \cr
   &\approx& (1-b) Br(B\to DX)\,Br(\overline B\to DX)
   +{b\over2}\,Br(B\to DX)\,Br(B\to DX)\,,
\end{eqnarray}
and
\begin{eqnarray}\label{ndds}
   n(DD^+_s) &=& (1-b)\,\left[ Br(B\to DX)\,Br(\overline B\to D^+_sX)
   +Br(B\to D^+_sX)\,Br(\overline B\to DX)\right] \cr
   &&+{b\over2}\left[Br(B\to DX)\,Br(B\to D^+_sX)+
   Br(\overline B\to DX)\,Br(\overline B\to D^+_sX)\right] \cr 
   &\approx& (1-b)\,Br(B\to DX)\,Br(\overline B\to D^+_sX)\,.
\end{eqnarray}   
where $b\approx0.076$ is the $B-\overline B$ mixing parameter~\cite{HS}.
It is defined as the fraction of $B$ meson events which are $BB$ or
$\overline B\,\overline B$, and is measured directly from lepton-lepton
sign correlations,
\begin{equation}
   b = {N_{BB}+N_{\overline B\,\overline B}\over
        N_{B\overline B}+N_{BB}+N_{\overline B\,\overline B}}
     = {N_{\ell^+\ell^+}+N_{\ell^-\ell^-}\over
        N_{\ell^+\ell^-}+N_{\ell^+\ell^+}+N_{\ell^-\ell^-}}\,.
\end{equation}
Combining Eqs.~(\ref{dmulti}) and (\ref{ndd}), we may extract
$Br(B\to DX)$ and $Br(B\to\overline DX)$ separately, if we neglect
$CP$ violation and impose the constraint $Br(\overline B\to DX) =
Br(B\to\overline DX)$.  Analogously, we may extract $Br(B\to
D^+_sX)$ and $Br(B\to D^-_sX)$.  Another method for determining individual
branching ratios would involve tagging the flavor of the $B$ which
produced the charmed hadron by measuring the charge of a hard primary
lepton from the other $B$ in the event.

Invoking a large violation of local duality has some implications for the
pattern of $B$ meson decays which may be different from what would be
expected if local duality held and an unusually small value of $m_c$
were used to explain the measured $B$ semileptonic branching ratio. For
example, a violation of local duality in the $b\to c\bar cs$ channel
could lead to quite different lifetimes for the $B$, $B_s$ and
$\Lambda_b$, differences which are small in the operator product
expansion because they arise only from higher dimension operators.
However, since the effective Hamiltonian for this process has isospin
zero, the equality of the $B^0$ and $B^-$ lifetimes would not be
disturbed.  Similarly, violations of local duality in $b\to c\bar ud'$
could lead to unequal $B^0$ and $B^-$ lifetimes. $B$ decay event shapes
can also provide a test of the free quark decay picture for the
$b\to c \bar ud'$ decay channel~\cite{LSW}.

\section{Concluding remarks}

We have examined whether the measured $B$ meson semileptonic branching
ratio can be explained within the conventional application of the
operator product expansion, in which operators of low dimension are kept
and perturbative corrections are included to a few orders in $\alpha_s$. 
We have found that this scenario would require an unusually small value
for $m_c$. If instead the explanation lies outside the conventional
application of the operator product expansion, then a failure of local
duality in the $b\to c\bar cs'$ channel is the likely explanation for the
discrepancy with experiment. In either case, we expect the number of
charmed hadrons per $B$ decay to be approximately 1.3.  Unfortunately,
the present data on charm multiplicities do not support such a large
value of $n_c$. From our perspective, the unusual feature of inclusive
$B$ decay is not the semileptonic branching ratio alone, nor the charm
multiplicity alone, but rather the combination of the two. Together, they
would seem to suggest a significant violation of local duality in the
$b\to c\bar ud'$ nonleptonic decay process.  From a theoretical point of view,
however, such a resolution would be somewhat unsettling, as it would
indicate a breakdown in the computation of the nonleptonic decay rate in
the region where it is expected to be the most reliable; we understand
why such a conclusion was resisted by the authors of Ref.~\cite{SVbaff}. 
Still, it remains an open possibility, indicating perhaps that the
invocation of local duality in quark decay requires a considerably larger
energy release than has been na\"{\i}vely hoped or expected.  Given the
apparent difficulties in performing a reliable computation of the
nonleptonic decay rate, then, the CKM matrix element
$V_{cb}$ should be extracted from the $B$ semileptonic decay width rather
than from the $B$ lifetime. The uncertainties in such an extraction arise
primarily from the choice of $m_b$ and subtraction point $\mu$, and are
discussed in detail in Refs.~\cite{LS,SUV}.

\acknowledgements
After this paper was completed, we learned that the order $\alpha_s$
radiative correction to the semileptonic decay width has been calculated
analytically, including all dependence on $m_c/m_b$, by Y.~Nir\cite{nir}.
We are indebted to B.~Blok, T.E.~Browder, P.S.~Cooper, B.~Grinstein,  
J.D.~Lewis, M.~Luke, Y.~Nir, M.~Savage, M.~Shifman, N.G.~Uraltsev and
A.I.~Vainshtein for useful conversations.  A.F.~and I.D.~would like to
thank the Institute for Theoretical Physics, where this work was
initiated, for their gracious hospitality.  This work was supported by
the Department of Energy under Grants  DOE-FG03-90ER40546,
DE-AC03-81ER40050 and DE-AC02-76CHO3000, and by the National Science
Foundation under Grant PHY89-04035.

\begin{figure}
\caption{Contours in the complex $v\cdot\hat q$ plane, for fixed  
$\hat q^2$ and $y$.  The gap between the cuts extends for $(1+\hat
q^2-\hat m_q^2)/2 < v\cdot\hat q < ((2+\hat m_q)^2-\hat q^2-1)/2$.  The
endpoints of the contour integral are at $v\cdot\hat q=(y+\hat
q^2/y)/2\pm i\epsilon$.}
\label{vdotqfig}
\end{figure}

\begin{figure}
\caption{The phase space suppression factor $G(\hat m_c)$, as an  
implicit function of $m_b$ with $m_b-m_c=3.34\,\text{GeV}$ held fixed.}
\label{Gfig}
\end{figure}

\begin{figure}
\caption{The radiative correction $P(\mu)$.  The upper curve  
corresponds to $\Lambda^{(5)}_{\overline{\text{MS}}}=220\,\text{MeV}$, the
middle curve to $\Lambda^{(5)}_{\overline{\text{MS}}}=180\,\text{MeV}$,
and the lower curve to
$\Lambda^{(5)}_{\overline{\text{MS}}}=140\,\text{MeV}$.  We take
$m_b=4.8\,\text{GeV}$.}
\label{Pfig}
\end{figure}

\end{document}